\documentclass{PoS}

\title{How far can you go ? Surprises and pitfalls
in three-flavour chiral extrapolations}

\ShortTitle{Surprises and pitfalls
in three-flavour chiral extrapolations}

\author{\speaker{S\'ebastien Descotes-Genon}\thanks{Work supported in part by the EU Contract No. MRTN-CT-2006-035482, \lq\lq
FLAVIAnet''.}\\
        Laboratoire de Physique Th\'eorique,\\ 
        CNRS/Univ. Paris-Sud 11 (UMR 8627),
        91405 Orsay Cedex, France\\
        E-mail: \email{descotes@th.u-psud.fr}}

\abstract{

The increasing accuracy of experimental data in flavour physics requires
a corresponding improvement on the theoretical side, in particular
concerning the non-perturbative dynamics of QCD. This has prompted the lattice
community to aim at an unprecedented accuracy in form factors and matrix
elements. However,
in the light sector, the meson masses remain too heavy for an interpolation,
which makes it necessary to rely on Chiral Perturbation Theory to 
perform extrapolations in the light quark masses. This makes it all the more necessary to
assess precisely the range of validity of this theory.
More precisely, the presence of
strange quark pairs in the sea may have a significant impact
of the pattern of chiral symmetry breaking : in particular 
large differences can occur between the chiral limits of 
two and three massless flavours (i.e., whether $m_s$ is kept at its physical
value or sent to zero). We recall some indications of such a
scenario in QCD, in relation with the peculiar dynamics of the scalar sector.
We explain how this could affect the convergence of three-flavour chiral
series, commonly used to extrapolate the results of lattice
simulations. Finally, we indicate how lattice simulations 
with three dynamical flavours could unveil such an effect
through the quark-mass dependence of light meson masses and
decay constants.}

\FullConference{The XXV International Symposium on Lattice Field Theory\\
                 July 30-4 August 2007\\
                 Regensburg, Germany}

\begin{document}

\section{Two chiral limits of interest}\label{sec:msdep}

Because of the mass hierarchy among light quarks,
the strange quark may play a special role in the low-energy dynamics of
QCD. It is light enough to allow for a combined expansion of
observables in powers of $m_u,m_d,m_s$ around the $N_f=3$
chiral limit (meaning 3 massless flavours): $m_u=m_d=m_s=0$.
But it is sufficiently heavy to induce significant 
changes in order parameters from the $N_f=3$ chiral limit to
the $N_f=2$ chiral limit: $m_u=m_d=0$ and $m_s$ physical.
Finally, it is too light to suppress efficiently
loop effects of massive $\bar{s}s$ pairs (contrary to heavier quarks)~\cite{param}.

Two different versions of Chiral Perturbation Theory have been developed
around these two limits. In the $N_f=2$ massless limit, only the pions play
a particular role and thus are the only available degrees of freedom. 
The $N_f=3$ chiral limit promotes pions, kaons and $\eta$ as the degrees 
of freedom : this second version of $\chi$PT is richer, discusses 
more processes in a larger range of energy, but contains more 
unknown low-energy constants (LECs) and may have a slower convergence. 
In each limit, the LECs encode the pattern of chiral symmetry breaking, cannot
be computed within $\chi$PT and must be determined from experiment. Obviously,
LECs in both theories are related since the $N_f=2$ theory can be obtained
from $N_f=3$ $\chi$PT by restricting it to the pion sector and 
integrating out kaons and eta, treated as massive particles. 
However, the details of the connection 
between the two theories remain under debate.

Indeed, due to $\bar{s}s$ sea-pairs, order parameters
such as the quark condensate and the pseudoscalar decay constant,
$\Sigma(N_f)=-\lim_{N_f} \langle\bar{u}u\rangle$ and 
$F^2(N_f)=\lim_{N_f} F^2_\pi$,
can reach significantly different values in the two chiral limits 
($\lim_{N_f}$ denoting the chiral limit with $N_f$ massless 
flavours)~\cite{param}.
An illustration is provided by the quark condensate in the two limits:
\begin{eqnarray}
\Sigma(2)&=&\Sigma(2;m_s)=\Sigma(2;0)
   + m_s \frac{\partial\Sigma(2;m_s)}{\partial m_s} + O(m_s^2)\\
&=&\Sigma(3)
   + m_s \lim_{m_u,m_d\to 0}
      i\int d^4x\ \langle 0|\bar{u}u(x)\, \bar{s}s(0) |0\rangle + O(m_s^2) \label{eq:sigma}
\end{eqnarray}
Here, $\bar{s}s$-pairs are involved through the two-point correlator 
$\langle (\bar{u}u) (\bar{s}s)\rangle$, which violates the Zweig rule in the vacuum
(scalar) channel~\footnote{Like all the terms in the $m_s$-expansion of
  $\Sigma(2)$, this correlator is a $SU_L(2)\times SU_R(2)$ order parameter related to the 
spontaneous breakdown of $N_f=2$ chiral 
symmetry, and this for any value of $m_s$. The (scheme-dependent) high-energy
contributions to eq.~(\ref{eq:sigma}) should be proportional to $m=m_u=m_d$ and thus drop from
the relation once the $N_f=2$ chiral limit has been taken (in contrast with
the quark condensates arising in OPE at non-vanishing $m_u,m_d,m_s$, which
exhibit such ultraviolet divergences).}.
This loop effect disappears in the large-$N_c$ limit, which
is known to fail in the $0^{++}$ channel due to the complicated structure of
broad resonances, corresponding to poles of the scattering matrix 
located far away from the real axis (see for instance~\cite{reson} for a
discussion  of the lightest scalar resonances). 

Arguments based on the spectrum of the Dirac operator~\cite{param} indicate that this effect should
suppress order parameters when $m_s\to 0$: 
$\Sigma(2) \geq \Sigma(3)$ and 
$F^2(2) \geq  F^2(3)$. Since the quark condensate(s)
and the pseudoscalar decay constant(s) are the building blocks
of the two versions of $\chi$PT at leading order,  
a strong decrease from
$N_f=2$ to $N_f=3$ should have a direct impact on the 
structure of the two theories. We discuss some available data on 
the pattern of chiral symmetry breaking in both chiral limits in turn.

\section{The situation in the $N_f=2$ limit}

A few years ago, the Brookhaven E865 collaboration 
provided new data on $K_{\ell 4}$ decays~\cite{E865}.
Building upon the dispersive analysis
of $\pi\pi$ scattering~\cite{royeq}, we extracted
the two-flavour order parameters~\cite{pipi}:
\begin{equation}
X(2)=\frac{(m_u+m_d)\Sigma(2)}{F_\pi^2M_\pi^2}=0.81\pm 0.07 \qquad
Z(2)=\frac{F^2(2)}{F_\pi^2}=0.89\pm 0.03
\end{equation}
A different analysis, with 
an additional theoretical input from the scalar radius of the pion, led to a larger 
value for $X(2)$~\cite{CGL}. 

The situation is somewhat modified by new data from the NA48
collaboration~\cite{NA48}, which show some discrepancy with the E865 phase
shifts in the higher end of the allowed phase space. 
The role of isospin breaking corrections is under discussion currently.
The preliminary
 values of the phase shifts~\cite{NA48} tend to increase the value
of the $I=J=0$ $\pi\pi$ scattering length, and to decrease the
value of the two-flavour quark condensate, possibly pushing $X(2)$ down to 0.7.

The Gell-Mann--Oakes--Renner relation and its siblings correspond to
$X(2)=Z(2)=1$ and such deviations from unity may seem fairly unimpressive. However one should remember
these are $N_f=2$ chiral expansions in powers of $m_u$ and $m_d$ \emph{only}.
Indeed, $X(2)$ and $Z(2)$ indicate the convergence of $N_f=2$ chiral expansions
of $F_\pi^2 M_\pi^2$ and $F_\pi^2$ respectively : they measure the relative size
of the leading-order term in these expansions. One usually
expects a far quicker convergence of $N_f=2$ chiral series, with
much smaller next-to-leading order corrections (below 10\%)~\cite{pipi}.

\section{The situation in the $N_f=3$ limit}

To include $K$- and $\eta$-mesons dynamically, one must use
three-flavour $\chi$PT,
where the expansion in the three light quark masses starts 
around the $N_f=3$ vacuum $m_u,m_d,m_s=0$.
Here, strange sea-quark loops
may have a dramatic effect on $N_f=3$ chiral expansions.
The leading-order (LO) term, which depends on the $O(p^2)$ low-energy
constants $F^2(3)$ and $\Sigma(3)$, would be damped, whereas
next-to-leading-order (NLO) corrections could be enhanced, in
particular those related to the violation of the Zweig rule in the scalar sector.
For instance, the Gell-Mann--Oakes--Renner relation would not be
saturated by its LO term and would receive sizeable numerical
contributions from terms treated as NLO in the chiral counting scheme.

Indirect estimates~\cite{uuss,pika} suggest a very
significant effect. At NLO the violation of the Zweig rule in the 
scalar sector is encoded in the two LECs $L_4$ and $L_6$. Dispersive
estimates of the correlator $\langle (\bar{u}u) (\bar{s}s) \rangle$,
related to $L_6$~\cite{uuss}, indicate that the quark condensate
may drop by a half from $N_f=2$ to $N_f=3$, i.e. when $m_s$ is sent
from its physical value down to zero (such a decrease has been observed by
the MILC collaboration~\cite{MILC}).
On the other hand~\cite{pika}, 
the dispersive study of low-energy $\pi K$ scattering
through Roy-Steiner equations and the naive comparison with NLO $N_f=3$ 
chiral expansions yields $L_4(M_\rho)=(0.53 \pm 0.39)\cdot 10^{-3}$. 

Such
a value may seem rather small, but one should not forget the $m_s$-enhancement
of the contributions of $L_4$ and $L_6$ in chiral series. Take for instance
\begin{equation} \label{eq:fpi}
F_\pi^2 = F(3)^2 + 16(m_s+2m)B_0 \Delta L_4 + 16mB_0\Delta L_5 +O(m_q^2)
\end{equation}
where $B_0=-\lim_{m_u,m_d,m_s\to 0} \langle\bar{u}u\rangle/F_\pi^2$, and 
we have put together NLO low-energy constants and chiral logarithms :
$\Delta L_5=L_5(M_\rho)+0.67\cdot 10^{-3}$ and
$\Delta L_4=L_4(M_\rho)+0.51\cdot 10^{-3}$ (which is enhanced by a factor of
$m_s/m$). 
If we assume
that the LO contribution is numerically dominant (i.e., $F_\pi^2 = F(3)^2$ to
a very good approximation), we can perform the following manipulations:
 \begin{equation} \label{eq:fpinum}
\frac{F(3)^2}{F_\pi^2}=\frac{F^2(3)}{F^2(3)+O(m_q^2)}
  = 1- 8\frac{2M_K^2+M_\pi^2}{F_\pi^2}\Delta L_4
     - 8\frac{M_\pi^2}{F_\pi^2}\Delta L_5 + O(m_q^2)
  =1- 0.51  -0.04 + O(m_q^2)
\end{equation}
where we have used $1/(1+x)=1-x$ and eq.~(\ref{eq:fpi}) at the second step,
and the second and third terms of the last equality are obtained using
$L_4(M_\rho)=0.5\cdot 10^{-3}$ and $L_5(M_\rho)=1.4\cdot 10^{-3}$ respectively
(these values are used for illustrative purposes). Eq.~(\ref{eq:fpinum})
is clearly in contradiction with the original 
assumption $F_\pi^2 \simeq F(3)^2$. A similar game can be played with
$L_6$ and $F_\pi^2M_\pi^2$.

We can draw several conclusions from this
simple exercise. A small positive value of $L_4(M_\rho)$ or of $L_6(M_\rho)$ is enough to
invalidate the usual assumption of a rapid convergence of $N_f=3$ chiral
series (an issue to be remembered when one tries to extract the values of LECs
in lattice simulations). In addition, we may encounter chiral series 
$A=A_{LO} + A_{NLO} + A\delta A$ with a good overall convergence, i.e., $\delta A\ll 1$
but the numerical balance between LO and NLO depends on 
the relevance of strange sea-quark loops. 
The numerical competition between formal LO and NLO makes
approximations such as $1/(1+x)\simeq 1-x$ or 
$(m_s+m)B_0\simeq M_K^2$ rather hazardous, leading to potential 
contradictions and/or slow convergence for some chiral expansions. 

Actually, such difficulties in the convergence are encountered 
when NNLO computations are fitted to 
experimental data. A good example can be found among the reference fits
in ref.~\cite{bijnens}: for instance, the
so-called Fit 10 yields : $(M_\pi^2)_{th} = (M_\pi^2)_{exp} 
[0.736+0.006-0.258]$,
corresponding to the relative size of LO, NLO and NNLO respectively. If
confirmed, such difficulties in the convergence of three-flavour 
chiral expansions should be considered as a very serious problem and a source
of sizable systematics in lattice results relying on chiral extrapolations
on a large range of quark masses.

\section{Constraints from $\pi\pi$ and $\pi K$ scatterings}

We would like to cope with the potentially ``large'' values of 
$L_4$ and $L_6$, and the resulting numerical competition 
between formal LO and NLO contributions in chiral series.
To do so,
we have introduced a framework, called Resummed Chiral Perturbation
Theory (Re$\chi$PT)~\cite{rechpt}, where we
define the appropriate observables to consider and
the treatment of their chiral expansion~\cite{rechpt,resum,extrapol}:
\begin{enumerate}
\item Consider a subset of observables that are assumed to have a 
good overall convergence -- we call them ``good observables''. They
must form a linear space, which we choose to be that of 
connected QCD correlators
(of vector/axial currents and their divergences) as functions
of external momenta, away from any kinematic singularities. 
This rule selects in particular $F_P^2$ and $F_P^2M_P^2$ ($P=\pi,K$).
\item Take each observable and write its NLO chiral expansion in terms
of the chiral couplings : $F_0$, $B_0$, $L_i$\ldots
\item In theses formulae, reexpress the bare couplings 
in terms of physical quantities (masses, decay constants\ldots)
if justified by physics considerations (e.g., position of unitarity cuts)
The result is called ``bare expansion''.
\item In these bare expansions, reexpress $O(p^2)$ 
and $O(p^4)$ LECs in terms of:
\begin{equation} \label{eq:fundparam}
r=\frac{m_s}{m}\,, \qquad
X(3)=\frac{2m\Sigma(3)}{F_\pi^2M_\pi^2}\,, \qquad
Z(3)=\frac{F^2(3)}{F_\pi^2}\,,
\end{equation}
and NNLO remainders, using the bare expansions for 
the masses and decay constants of the pseudoscalar mesons $\pi,K,\eta$.
\end{enumerate}

This simple recipe provides a resummation of the potentially
large effect of the Zweig-rule violating couplings $L_4$ and
$L_6$~\cite{resum,rechpt}, hence the name of ``Resummed Chiral Perturbation
Theory'' (Re$\chi$PT) given to this framework. Since we want to cope with the
possibility of a numerical competition between (formal) LO and NLO terms in
chiral series, some usual $O(p^4)$ results are not valid any longer : for instance
$r=m_s/m$ is not fixed by $M_K^2/M_\pi^2$ and can vary from 8 to 40, $L_5$ is
not fixed by the ratio of $F_K/F_\pi$.

\begin{figure}[t]
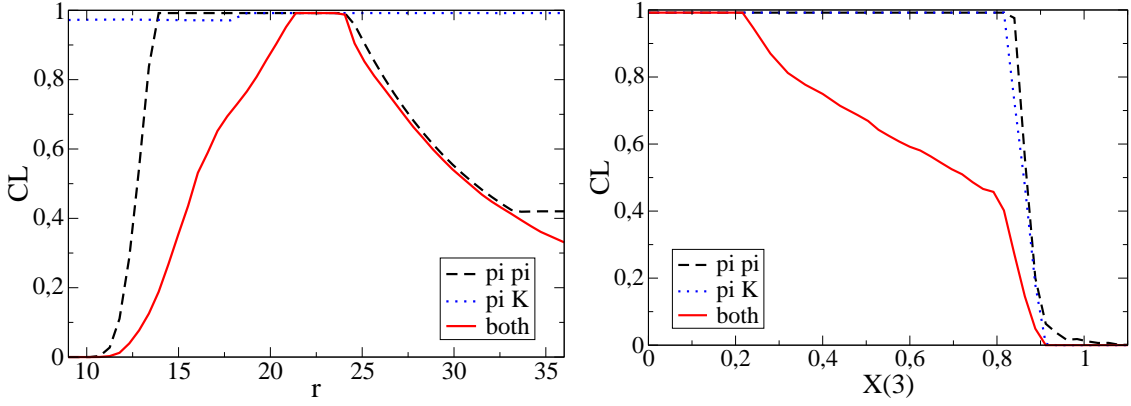

\includegraphics[height=5.3cm]{pr-1-col}
\hspace{0.1cm}
\includegraphics[height=5.3cm]{px-1-col}
\caption{Profiles for the confidence levels of $r=m_s/m$ (left) and
$X(3)=2m\Sigma(3)/(F_\pi^2M_\pi^2)$ (right). In each case, the results 
are obtained from $\pi\pi$ scattering only (dashed line), $\pi K$ scattering
only (dotted line), or both sources of information (solid line).}
\label{fig:CLs}
\end{figure}

In this framework, one can derive the NLO amplitudes corresponding
to $\pi\pi$ and $\pi K$ scatterings, which provide information on $N_f=2$ and
$N_f=3$ patterns of chiral symmetry breaking. The smallest uncertainties on the chiral
expansion are expected to occur in the unphysical region, far away from the
non-analyticities due to unitarity. One can exploit dispersive relations, such
as Roy equations~\cite{royeq} and Roy-Steiner equations~\cite{pika}, to reconstruct the amplitudes in
this unphysical point from the phase shifts from threshold up to energies around 1 GeV.
Matching the dispersive and chiral representations of the amplitude in a
frequentist framework provides constraints (in terms of confidence Levels) 
on the main parameters of interest for three-flavour $\chi$PT~\cite{rechpt}. 
As an
illustration, fig.~\ref{fig:CLs} shows the situation for $r=m_s/m$ and 
$X(3)=2m\Sigma(3)/(F_\pi^2M_\pi^2)$. A more detailed analysis provides the
following constraints from the combination of $\pi\pi$ and $\pi K$ scatterings:
\begin{equation}
r \geq 14.8\,, \quad  X(3)\leq 0.83\,, \quad Y(3)\leq 1.1\,, 
  \quad 0.18 \leq Z(3)\leq 1\,. \qquad  [68 \% {\rm CL}]
\end{equation}

Therefore, one can extract only mild constraints on the parameters of interest
from experimental data~: it is far from clear whether the usual description of
chiral symmetry breaking, triggered by a large quark condensation and
independent from strange sea-quark effects, holds or not.

\section{Three-flavour $\chi$PT and lattice simulations}

Lattice simulations may suffer from unexpected systematics in chiral
extrapolations
due to a strong $m_s$-dependence of chiral order parameters, and the 
resulting numerical competition between LO and NLO in three-flavour chiral series.
But they may also shed some light on this problem, since they allow one
to vary the quark masses, and thus enhance or suppress Zweig-rule violating
contributions accordingly~\cite{extrapol}.
We consider a lattice simulation with (2+1) flavours : 
two flavours are set to a common mass $\tilde{m}$, whereas the third one 
is kept at the same mass as the physical strange quark $m_s$.
Each quantity $X$ observed in the physical situation $(m,m,m_s)$ has
a lattice counterpart $\tilde{X}$ for $(\tilde{m},\tilde{m},m_s)$.
One can study the variation of
$\tilde{F}_P^2$ and $\tilde{F}_P^2\tilde{M}_P^2$ according to
\begin{equation} 
q=\frac{\tilde{m}}{m_s}\,, \qquad r=\frac{m_s}{m}\,, \qquad
X(3)=\frac{2m\Sigma(3)}{F_\pi^2M_\pi^2}\,, \qquad
Z(3)=\frac{F^2(3)}{F_\pi^2}
\end{equation}

\begin{figure}[t]
\includegraphics[width=7.5cm]{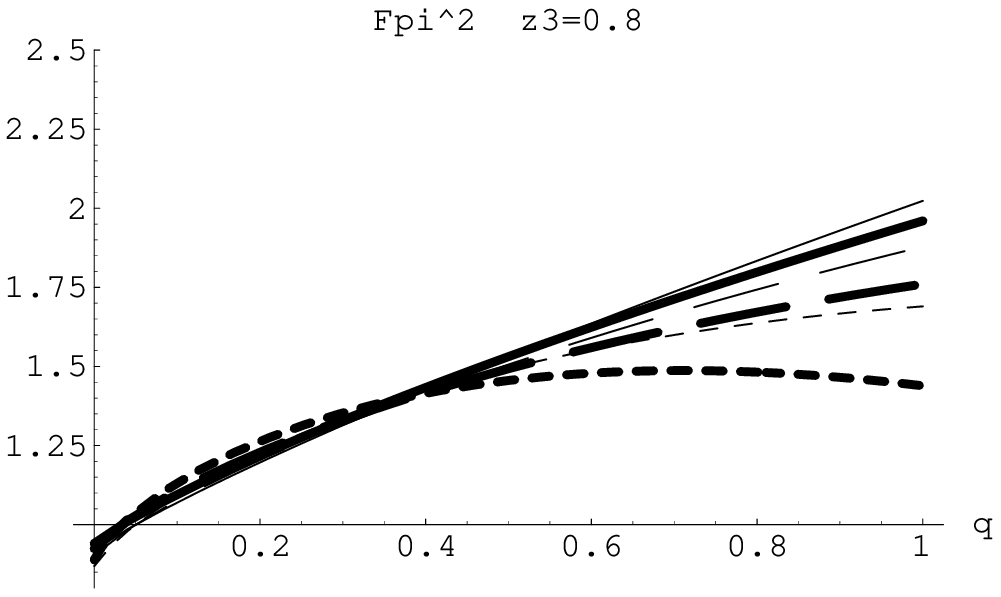}
\includegraphics[width=7.5cm]{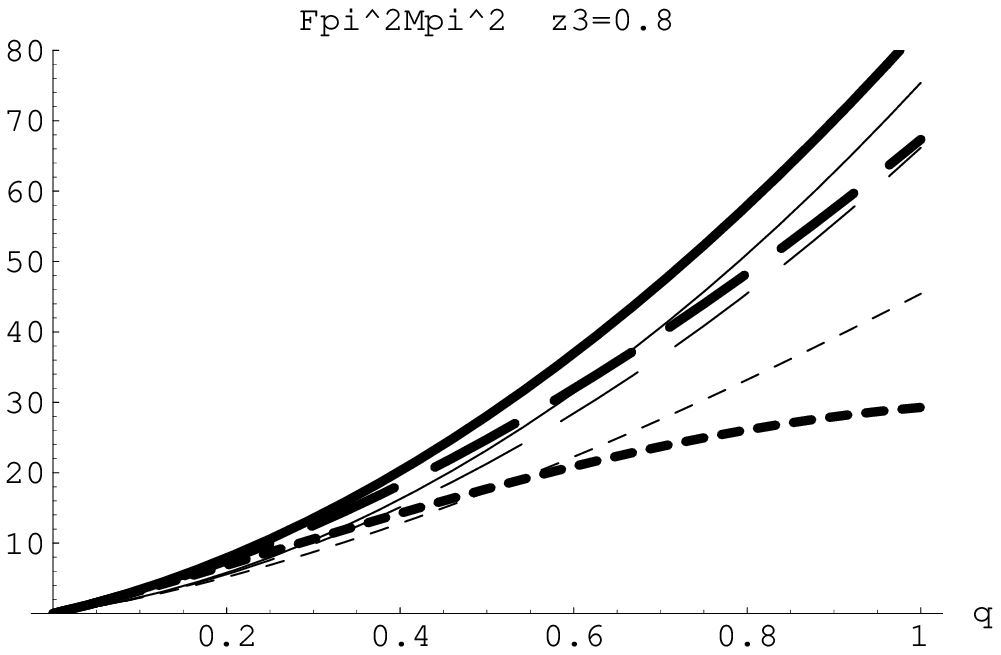}
\caption{$\tilde{F}_\pi^2/F_\pi^2$ (left) and
$\tilde{F}_\pi^2\tilde{M}_\pi^2/(F_\pi^2M_\pi^2)$
(right) as functions of $q=\tilde{m}/m_s$.
Solid, long-dashed and dashed curves correspond respectively
to $X(3)=0.8, 0.4, 0.2$. Thick (thin) lines are drawn for $r=30$ (20).
We set $Z(3)=0.8$ and NNLO remainders are neglected.}
\label{fig:inflim}
\end{figure}

Figure~\ref{fig:inflim} illustrates how
the dependence of hadronic observables on quark masses is related
to Zweig-rule violating sea-quark effects: if the latter
are large [$X(3)=0.2<X(2)$, dashed line], the curves bend more than in the case 
of negligible sea-quark loops [$X(3)=0.8 \sim X(2)$, solid line].

Actually, lattice simulations are performed in a finite spatial box, 
whereas time is sent to infinity to single out the ground state. 
For sufficiently large boxes, the low-energy effective theory 
is identical to $\chi$PT, with the same values of the 
chiral couplings as in an infinite volume. The only difference arises in the
propagators of the Goldstone modes, so that the chiral series
for $\tilde{F}_P^2$ and $\tilde{F}_P^2\tilde{M}_P^2$ change only through
tadpole logarithms. One can extend Re$\chi$PT to cope with such a
problem by noting that only a part of finite-volume
corrections computed at NLO in $\chi$PT has to involve the physical
masses of the Goldstone bosons~\cite{extrapol}.
The relative finite-volume
corrections for these observables are significant for $L=1.5$ fm, but
much smaller for $L=2.5$ fm. In addition, the corrections are smaller 
for $\tilde{F}_\pi^2\tilde{M}_\pi^2$ than for the decay constant. By inspection,
we see that $\tilde{F}_P^2\tilde{M}_P^2$ ($P=\pi,K$) in 
large volumes is a quantity for which
we manage a good control of finite-volume effects, whatever the impact
of sea $s\bar{s}$ pairs on the pattern of $N_f=3$ chiral symmetry
breaking~\cite{extrapol}. 
More specifically, the two
ratios:
\begin{equation}
R_\pi=\frac{1}{q}
  \frac{\tilde{F}_\pi^2 \tilde{M}_\pi^2}{F_\pi^2 M_\pi^2}\,,
\qquad \qquad
R_K=\frac{2}{(q+1)}\frac{\tilde{F}_K^2 \tilde{M}_K^2}{F_K^2 M_K^2}\,,
\end{equation}
are only mildly affected by finite-volume effects, 
and their $q$-dependence should provide valuable information
on the importance of strange-quark loops. 
NNLO remainders blur slightly the picture, but they do not prevent
the assessment of the effect.

\section{Conclusion}
The presence of massive $s\bar{s}$-pairs in the QCD vacuum may induce
significant differences in the pattern of chiral symmetry breaking
between the $N_f=2$ and $N_f=3$ chiral limits.
This effect, related to the violation of the Zweig rule in the scalar
sector, may destabilise three-flavour chiral expansions numerically and may yield
large systematics for chiral extrapolations of lattice data down to very light
quark masses. We have compared the situation in the two- and three-flavour
chiral limits : the combination of $\pi\pi$ and $\pi K$ data does not favour 
the usual picture of a large quark condensation independent of the number
of massless flavours. We indicated how chiral extrapolations of lattice results
with three dynamical quarks could be affected in the case of significant
differences between  the $N_f=2$ and $N_f=3$ chiral limits. In addition, we
suggested a way to test this scenario on the lattice, through
two dimensionless ratios which prove sensitive to this effect, 
with only a mild impact of finite-volume corrections.

\end{document}